\documentclass[12pt,a4paper]{article}

\usepackage{xcolor}

\usepackage{ifthen} 
\newboolean{pdflatex}
\setboolean{pdflatex}{true} 

\newboolean{articletitles}
\setboolean{articletitles}{true} 

\newboolean{uprightparticles}
\setboolean{uprightparticles}{false} 

\def\paperauthors{LHCb collaboration} 
\def\paperasciititle{Measurement of the branching fraction of Bs->K_S K_S decays} 
\def\papertitle{Measurement of the branching fraction of the decay \decay{\Bs}{\KS\KS}} 
\def\paperkeywords{{High Energy Physics}, {LHCb}} 
\def\papercopyright{\the\year\ CERN for the benefit of the LHCb collaboration} 
\def\paperlicence{CC-BY-4.0 licence}
\def\paperlicenceurl{https://creativecommons.org/licenses/by/4.0/}

\usepackage[top=1in, bottom=1.25in, left=1in, right=1in]{geometry}

\usepackage{microtype}
\usepackage{lineno}  
\usepackage{xspace} 
\usepackage{caption} 

\usepackage{graphicx}  
\usepackage{color}
\usepackage{colortbl}
\graphicspath{{./figs/}} 
\DeclareGraphicsExtensions{.pdf,.PDF,png,.PNG}

\usepackage{amsmath} 
\usepackage{amssymb}
\usepackage{amsfonts}
\usepackage{upgreek} 

\newcommand*\patchAmsMathEnvironmentForLineno[1]{%
	\expandafter\let\csname old#1\expandafter\endcsname\csname #1\endcsname
	\expandafter\let\csname oldend#1\expandafter\endcsname\csname
	end#1\endcsname
	\renewenvironment{#1}%
	{\linenomath\csname old#1\endcsname}%
	{\csname oldend#1\endcsname\endlinenomath}%
}
\newcommand*\patchBothAmsMathEnvironmentsForLineno[1]{%
	\patchAmsMathEnvironmentForLineno{#1}%
	\patchAmsMathEnvironmentForLineno{#1*}%
}
\AtBeginDocument{%
	\patchBothAmsMathEnvironmentsForLineno{equation}%
	\patchBothAmsMathEnvironmentsForLineno{align}%
	\patchBothAmsMathEnvironmentsForLineno{flalign}%
	\patchBothAmsMathEnvironmentsForLineno{alignat}%
	\patchBothAmsMathEnvironmentsForLineno{gather}%
	\patchBothAmsMathEnvironmentsForLineno{multline}%
	\patchBothAmsMathEnvironmentsForLineno{eqnarray}%
}

\usepackage{hyperxmp}

\usepackage[pdftex,
pdfauthor={\paperauthors},
pdftitle={\paperasciititle},
pdfkeywords={\paperkeywords},
pdfcopyright={Copyright (C) \papercopyright},
pdflicenseurl={\paperlicenceurl}]{hyperref}

\usepackage[colorinlistoftodos,textsize=scriptsize]{todonotes}

\usepackage[all]{hypcap} 

\usepackage{xspace} 
\usepackage{upgreek}

\newcommand{\offsetoverline}[2][0.1em]{\kern #1\overline{\kern -#1 #2}}%

\def\lhcb   {\mbox{LHCb}\xspace}

\def\belle  {\mbox{Belle}\xspace}

\def\velo   {VELO\xspace}

\def\MagUp {\mbox{\em Mag\kern -0.05em Up}\xspace}

\ifthenelse{\boolean{uprightparticles}}%
{

	\def\Ppi         {\ensuremath{\uppi}\xspace}

	\def\Pphi        {\ensuremath{\upphi}\xspace}

	\def\Ppsi        {\ensuremath{\uppsi}\xspace}

	\def\PDelta      {\ensuremath{\Delta}\xspace}                 
	\def\PXi         {\ensuremath{\Xi}\xspace}                 
	\def\PLambda     {\ensuremath{\Lambda}\xspace}                 
	\def\PSigma      {\ensuremath{\Sigma}\xspace}                 
	\def\POmega      {\ensuremath{\Omega}\xspace}                 
	\def\PUpsilon    {\ensuremath{\Upsilon}\xspace}

	\def\PB      {\ensuremath{\mathrm{B}}\xspace}                 
	                 
	\def\PD      {\ensuremath{\mathrm{D}}\xspace}

	\def\PK      {\ensuremath{\mathrm{K}}\xspace}

	\def\Pb      {\ensuremath{\mathrm{b}}\xspace}                 
	\def\Pc      {\ensuremath{\mathrm{c}}\xspace}                 
	\def\Pd      {\ensuremath{\mathrm{d}}\xspace}

	\def\Pi      {\ensuremath{\mathrm{i}}\xspace}

	\def\Pp      {\ensuremath{\mathrm{p}}\xspace}

	\def\Ps      {\ensuremath{\mathrm{s}}\xspace}

}
{

	\def\Ppi         {\ensuremath{\pi}\xspace}

	\def\Pphi        {\ensuremath{\phi}\xspace}

	\def\Ppsi        {\ensuremath{\psi}\xspace}                 
	                 
	\mathchardef\PDelta="7101
	\mathchardef\PXi="7104
	\mathchardef\PLambda="7103
	\mathchardef\PSigma="7106
	\mathchardef\POmega="710A
	\mathchardef\PUpsilon="7107
	                 
	\def\PB      {\ensuremath{B}\xspace}                 
	                 
	\def\PD      {\ensuremath{D}\xspace}

	\def\PK      {\ensuremath{K}\xspace}

	\def\Pb      {\ensuremath{b}\xspace}                 
	\def\Pc      {\ensuremath{c}\xspace}                 
	\def\Pd      {\ensuremath{d}\xspace}

	\def\Pi      {\ensuremath{i}\xspace}

	\def\Pp      {\ensuremath{p}\xspace}

	\def\Ps      {\ensuremath{s}\xspace}

}

\makeatletter
\ifcase \@ptsize \relax
\newcommand{\miniscule}{\@setfontsize\miniscule{4}{5}}
\or
\newcommand{\miniscule}{\@setfontsize\miniscule{5}{6}}
\or
\newcommand{\miniscule}{\@setfontsize\miniscule{5}{6}}
\fi
\makeatother

\DeclareRobustCommand{\optbar}[1]{\shortstack{{\miniscule (\rule[.5ex]{1.25em}{.18mm})}
		\\ [-.7ex] $#1$}}

\def\dquark    {{\ensuremath{\Pd}}\xspace}
\def\dquarkbar {{\ensuremath{\overline \dquark}}\xspace}

\def\squark    {{\ensuremath{\Ps}}\xspace}

\def\cquark    {{\ensuremath{\Pc}}\xspace}

\def\bquark    {{\ensuremath{\Pb}}\xspace}

\def\pion   {{\ensuremath{\Ppi}}\xspace}
\def\piz    {{\ensuremath{\pion^0}}\xspace}
\def\pip    {{\ensuremath{\pion^+}}\xspace}
\def\pim    {{\ensuremath{\pion^-}}\xspace}
\def\pipm   {{\ensuremath{\pion^\pm}}\xspace}
\def\pimp   {{\ensuremath{\pion^\mp}}\xspace}

\def\kaon    {{\ensuremath{\PK}}\xspace}
\def\Kbar    {{\kern 0.2em\overline{\kern -0.2em \PK}{}}\xspace}

\def\KorKbar {\kern 0.18em\optbar{\kern -0.18em K}{}\xspace}
\def\Kz      {{\ensuremath{\kaon^0}}\xspace}
\def\Kzb     {{\ensuremath{\Kbar{}^0}}\xspace}
\def\Kp      {{\ensuremath{\kaon^+}}\xspace}
\def\Km      {{\ensuremath{\kaon^-}}\xspace}
\def\Kpm     {{\ensuremath{\kaon^\pm}}\xspace}

\def\KS      {{\ensuremath{\kaon^0_{\mathrm{S}}}}\xspace}
\def\KL      {{\ensuremath{\kaon^0_{\mathrm{L}}}}\xspace}
\def\Kstarz  {{\ensuremath{\kaon^{*0}}}\xspace}
\def\Kstarzb {{\ensuremath{\Kbar{}^{*0}}}\xspace}

\def\Kstarp  {{\ensuremath{\kaon^{*+}}}\xspace}

\newcommand{\phiz}{\ensuremath{\Pphi}\xspace}

\def\Dbar    {{\kern 0.2em\overline{\kern -0.2em \PD}{}}\xspace}

\def\DorDbar {\kern 0.18em\optbar{\kern -0.18em D}{}\xspace}

\def\B       {{\ensuremath{\PB}}\xspace}
\def\Bbar    {{\ensuremath{\kern 0.18em\overline{\kern -0.18em \PB}{}}}\xspace}

\def\BorBbar    {\kern 0.18em\optbar{\kern -0.18em B}{}\xspace}
\def\Bz      {{\ensuremath{\B^0}}\xspace}

\def\Bu      {{\ensuremath{\B^+}}\xspace}

\def\Bp      {{\ensuremath{\Bu}}\xspace}

\def\Bs      {{\ensuremath{\B^0_\squark}}\xspace}
\def\Bsb     {{\ensuremath{\Bbar{}^0_\squark}}\xspace}
\def\BdorBs  {{\ensuremath{\B^0_{(\squark)}}}\xspace}

\def\Bds     {{\ensuremath{\B_{(\squark)}^0}}\xspace}

\def\psitwos  {{\ensuremath{\Ppsi{(2S)}}}\xspace}

\def\Y#1S{\ensuremath{\PUpsilon{(#1S)}}\xspace}

\def\proton      {{\ensuremath{\Pp}}\xspace}

\def\Lz          {{\ensuremath{\PLambda}}\xspace}

\def\LorLbar     {\kern 0.18em\optbar{\kern -0.18em \PLambda}{}\xspace}

\def\Lb           {{\ensuremath{\Lz^0_\bquark}}\xspace}

\def\BF         {{\ensuremath{\mathcal{B}}}\xspace}
\def\BR         {\BF}

\newcommand{\decay}[2]{\mbox{\ensuremath{#1\!\to #2}}\xspace}         

\def\to                 {\ensuremath{\rightarrow}\xspace}

\def\CP                {{\ensuremath{C\!P}}\xspace}

\newcommand{\phis}{{\ensuremath{\phi_{\squark}}}\xspace}

\def\AT#1     {\ensuremath{A_{\mathrm{T}}^{#1}}\xspace}           

\def\C#1      {\ensuremath{\mathcal{C}_{#1}}\xspace}                       
\def\Cp#1     {\ensuremath{\mathcal{C}_{#1}^{'}}\xspace}                    
\def\Ceff#1   {\ensuremath{\mathcal{C}_{#1}^{\mathrm{(eff)}}}\xspace}        
\def\Cpeff#1  {\ensuremath{\mathcal{C}_{#1}^{'\mathrm{(eff)}}}\xspace}       
\def\Ope#1    {\ensuremath{\mathcal{O}_{#1}}\xspace}                       
\def\Opep#1   {\ensuremath{\mathcal{O}_{#1}^{'}}\xspace}                    

\newcommand{\nospaceunit}[1]{\ensuremath{\text{#1}}}       
\newcommand{\aunit}[1]{\ensuremath{\text{\,#1}}}       

\newcommand{\tev}{\aunit{Te\kern -0.1em V}\xspace}
\newcommand{\gev}{\aunit{Ge\kern -0.1em V}\xspace}
\newcommand{\mev}{\aunit{Me\kern -0.1em V}\xspace}
\newcommand{\kev}{\aunit{ke\kern -0.1em V}\xspace}
\newcommand{\ev}{\aunit{e\kern -0.1em V}\xspace}
\newcommand{\mevc}{\ensuremath{\aunit{Me\kern -0.1em V\!/}c}\xspace}
\newcommand{\gevc}{\ensuremath{\aunit{Ge\kern -0.1em V\!/}c}\xspace}
\newcommand{\mevcc}{\ensuremath{\aunit{Me\kern -0.1em V\!/}c^2}\xspace}
\newcommand{\gevcc}{\ensuremath{\aunit{Ge\kern -0.1em V\!/}c^2}\xspace}

\def\mum  {\ensuremath{\,\upmu\nospaceunit{m}}\xspace}

\def\fb   {\ensuremath{\aunit{fb}}\xspace}
\def\invfb   {\ensuremath{\fb^{-1}}\xspace}

\newcommand{\chisq}{\ensuremath{\chi^2}\xspace}

\newcommand{\chisqip}{\ensuremath{\chi^2_{\text{IP}}}\xspace}

\def\gsim{{~\raise.15em\hbox{$>$}\kern-.85em
		\lower.35em\hbox{$\sim$}~}\xspace}
\def\lsim{{~\raise.15em\hbox{$<$}\kern-.85em
		\lower.35em\hbox{$\sim$}~}\xspace}

\def\pt         {\ensuremath{p_{\mathrm{T}}}\xspace}

\def\ptot       {\ensuremath{p}\xspace}

\def\evtgen     {\mbox{\textsc{EvtGen}}\xspace}

\def\geant      {\mbox{\textsc{Geant4}}\xspace}

\def\photos     {\mbox{\textsc{Photos}}\xspace}

\def\pythia     {\mbox{\textsc{Pythia}}\xspace}

\xspace

\def\tell1  {TELL1\xspace}
\def\ukl1   {UKL1\xspace}

\usepackage{xspace}
\usepackage{upgreek}

\def\BzToKSKS  {\decay{\Bz}{\KS\KS}}
\def\BsToKSKS  {\decay{\Bs}{\KS\KS}}
\def\BszToKSKS {\decay{\BdorBs}{\KS\KS}}

\def\BzToKzKzb  {\decay{\Bz}{\Kz\Kzb}}
\def\BsToKzKzb  {\decay{\Bs}{\Kz\Kzb}}
\def\BszToKzKzb {\decay{\BdorBs}{\Kz\Kzb}}

\def\BzToKSphi  {\decay{\Bz}{\phi\KS}}
\def\BzToKzphi  {\decay{\Bz}{\phiz\Kz}}

\def\Kzboptstar {{\ensuremath{\Kbar{}^{(*)0}}}\xspace}

\def\LbTopKSpi {\decay{\Lb}{\proton\KS\pim}}

\def\KSTopippim {\decay{\KS}{\pip\pim}}
\def\phiToKK {\decay{\phiz}{\Kp\Km}}

\newcommand{\fracs}{\ensuremath{f_{s}}\xspace}
\newcommand{\fracd}{\ensuremath{f_{d}}\xspace}

\newcommand{\mymev}{{\ensuremath{\mathrm{\,Me\kern -0.1em V}}}\xspace}
\newcommand{\mygev}{{\ensuremath{\mathrm{\,Ge\kern -0.1em V}}}\xspace}
\newcommand{\mytev}{{\ensuremath{\mathrm{\,Te\kern -0.1em V}}}\xspace}

\newcommand{\splot}{\ensuremath{{}_s\mathcal{P}\text{lot}}\xspace}

\def\Opeup#1#2   {\ensuremath{\mathcal{O}_{#1}^{#2}}\xspace}                       

\usepackage{cite} 
\usepackage{mciteplus}

\usepackage{longtable} 

\usepackage{siunitx} 
\usepackage{xfrac} 

\usepackage{cleveref} 
  \crefname{chapter}{Ch.\@}{Chs.\@}
  \crefname{section}{Sec.\@}{Secs.\@}
  \crefname{subsection}{Sec.\@}{Secs.\@}
  \crefname{figure}{Fig.\@}{Figs.\@}
  \crefname{table}{Table}{Tables} 
  \crefname{equation}{Eq.\@}{Eqs.\@}
  \crefname{appch}{Appendix}{Appendices}
  \crefname{appsec}{Appendix}{Appendices}
  \crefname{appsubsec}{Appendix}{Appendices}
  \crefname{appsubsubsec}{Appendix}{Appendices}

\usepackage{booktabs} 
\usepackage{tabularx}

\usepackage{float} 
\begin{document}

\renewcommand{\thefootnote}{\fnsymbol{footnote}}
\setcounter{footnote}{1}

\begin{titlepage}
	\pagenumbering{roman}
	
	\vspace*{-1.5cm}
	\centerline{\large EUROPEAN ORGANIZATION FOR NUCLEAR RESEARCH (CERN)}
	\vspace*{1.5cm}
	\noindent
	\begin{tabular*}{\linewidth}{lc@{\extracolsep{\fill}}r@{\extracolsep{0pt}}}
		\ifthenelse{\boolean{pdflatex}}
		{\vspace*{-1.5cm}\mbox{\!\!\!\includegraphics[width=.14\textwidth]{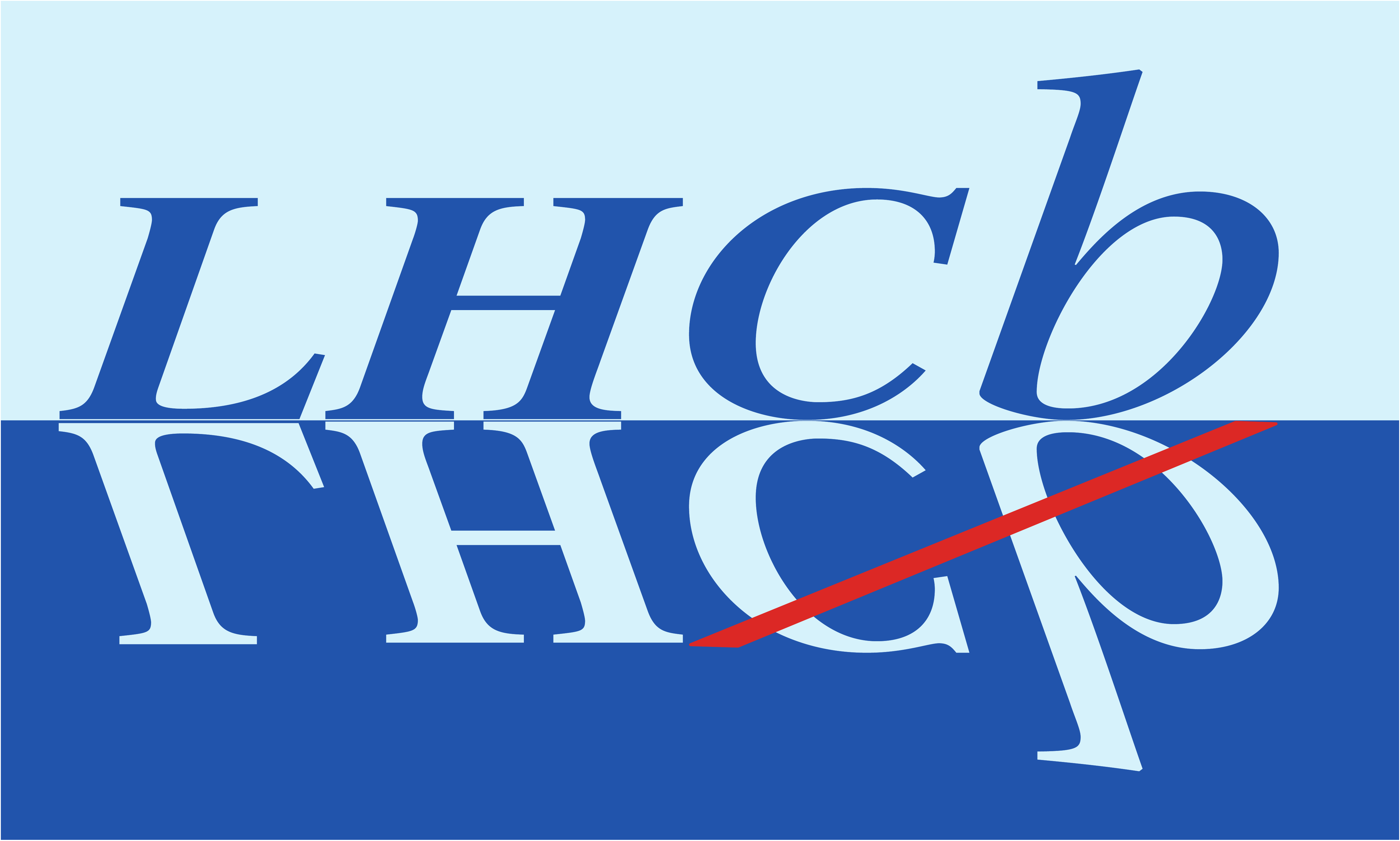}} & &}%
		{\vspace*{-1.2cm}\mbox{\!\!\!\includegraphics[width=.12\textwidth]{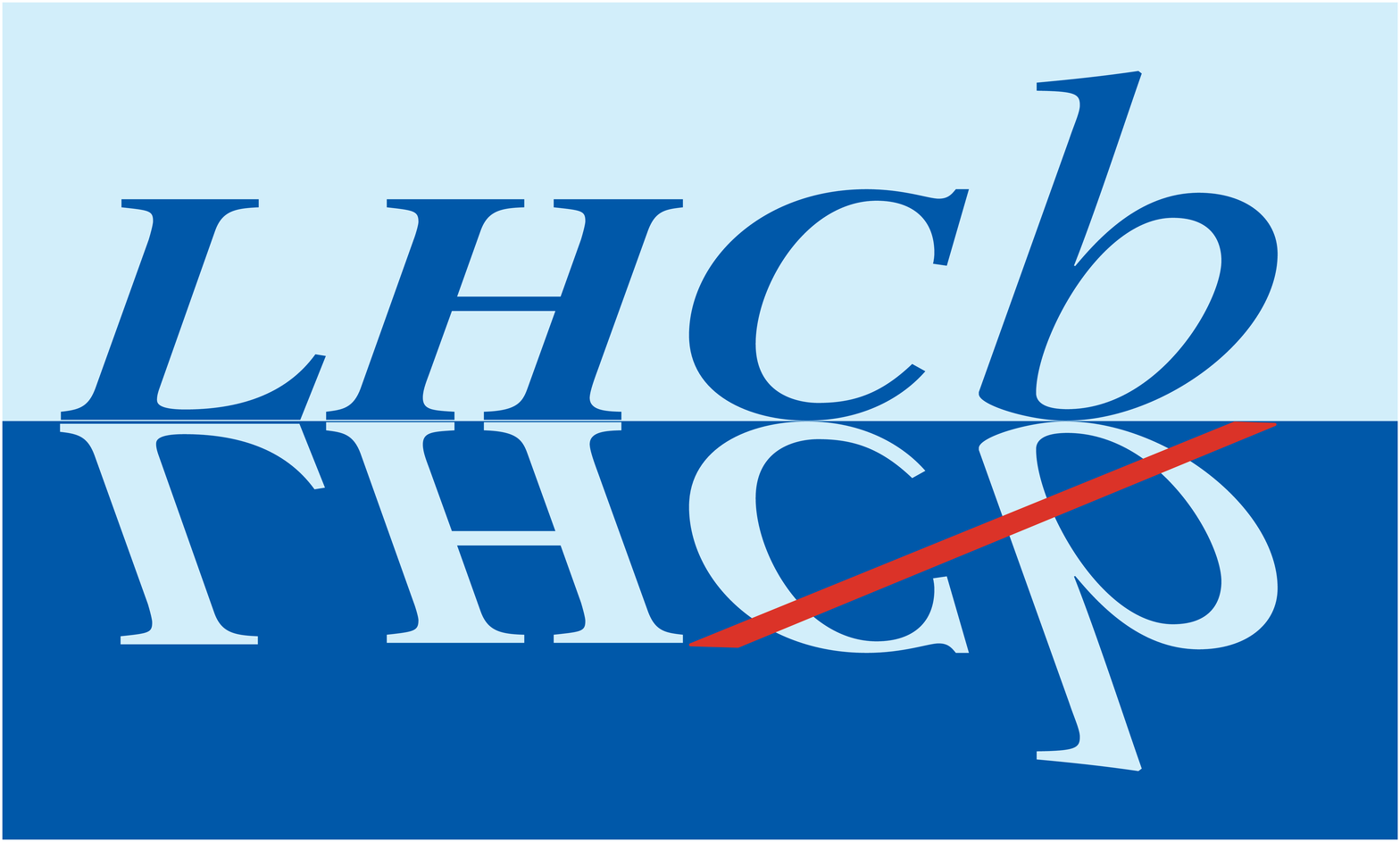}} & &}%
		\\
		& & CERN-EP-2020-008 \\  
		& & LHCb-PAPER-2019-030 \\  
		& & 12 August 2020 \\ 
		& & \\
	\end{tabular*}
	
	\vspace*{3.5cm}
	
	{\normalfont\bfseries\boldmath\huge
		\begin{center}
			\papertitle 
		\end{center}
	}
	
	\vspace*{2.0cm}
	
	\begin{center}
		\paperauthors\footnote{Authors are listed at the end of this paper.}
	\end{center}
	
	\vspace{\fill}
	
	\begin{abstract}
		\noindent
		A measurement of the branching fraction of the decay \BsToKSKS is performed using proton--proton collision data
		corresponding to an integrated luminosity of $\SI{5}{\invfb}$ collected by the \lhcb experiment between 2011 and 2016. The branching fraction is
		determined to be
		\begin{equation*}
		\BR(\BsToKSKS) = [8.3 \pm 1.6 \,(\text{stat}) \pm 0.9 \,(\text{syst}) 
		\pm 0.8 \,(\text{norm}) \pm 0.3 \,(\sfrac{f_s}{f_d})] \times 10^{-6} \,,
		\end{equation*}
		where the first uncertainty is statistical, the second is systematic, and the third and fourth are due to uncertainties on the branching fraction of the normalization mode \BzToKSphi
		and the ratio of hadronization fractions $\sfrac{f_s}{f_d}$. This is the most precise measurement of this branching fraction to date.
		Furthermore, a measurement of the branching fraction of the decay \BzToKSKS is performed relative to that of the \BsToKSKS channel, and is found to be
		\begin{equation*}
		\frac{\BR(\BzToKSKS)}{\BR(\BsToKSKS)} = [7.5 \pm 3.1 \,(\text{stat}) \pm 0.5 \,(\text{syst}) \pm 0.3 \,(\sfrac{f_s}{f_d})] \times 10^{-2} \,.
		\end{equation*}  
	\end{abstract}
	
	\vspace*{1.5cm}
	
	\begin{center}
		Published in Phys. Rev. D 102 (2020) 012011 
	\end{center}
	
	\vspace{\fill}
	
	{\footnotesize 
		\centerline{\copyright~\papercopyright. \href{\paperlicenceurl}{\paperlicence}.}}
	\vspace*{2mm}
	
\end{titlepage}


\newpage
\setcounter{page}{2}
\mbox{~}

\cleardoublepage

\renewcommand{\thefootnote}{\arabic{footnote}}
\setcounter{footnote}{0}


\pagestyle{plain} 
\setcounter{page}{1}
\pagenumbering{arabic}

\section{Introduction}

Flavour-changing neutral current processes, especially neutral $B$ meson decays to kaons and excited kaons, can be used as probes of the Standard Model and of the CKM unitarity triangle angle $\beta_{(s)}$. While decays such as ${\decay{B^0_{(s)}}{\Kstarz \Kstarzb}}$, ${\decay{B^0_s}{\Kstarz \Kzb}}$, and ${\decay{\Bz}{\Kp\Km}}$ have already been measured at the LHC~\cite{LHCb-PAPER-2019-004,LHCb-PAPER-2017-048,LHCB-PAPER-2018-045,LHCb-PAPER-2016-036}, decays of \bquark hadrons to final states containing only long-lived particles, such as \KS mesons or \Lz baryons, have never before been reported in a hadronic production environment. A measurement of the branching fraction of \BsToKzKzb decays can be used as input to future SM predictions, and is a first step toward a time-dependent measurement of \CP violation in this channel using future LHC data.

In the Standard Model, the decay amplitude of \BsToKzKzb is dominated by $\bquark \to \squark \dquarkbar \dquark$ loop transitions with gluon radiation, while other contributions, including color singlet exchange, are suppressed to the level of \mbox{$ \SI{5}{\percent}$}~\cite{Baek:2006pb} in the decay amplitude.
Predictions of this branching fraction within the SM lie in the range $(15-25)\times 10^{-6}$~\cite{Williamson:2006hb,Beneke:2003zv,PhysRevD.76.074018,PhysRevD.89.074046}, with calculations relying on a variety of theoretical approaches such as soft collinear effective theory, QCD factorisation, and perturbative leading-order and next-to-leading-order QCD. Beyond the Standard Model, possible contributions from new particles or couplings~\cite{Chang:2013hba,DescotesGenon:2006wc,Baek:2006pb,Ciuchini:2007hx,BHATTACHARYA2012403} can be probed by improved experimental precision on the branching fraction measurement.

The decay \BsToKzKzb was first observed by the \belle collaboration in 2016~\cite{PhysRevLett.116.161801}.
The branching fraction was determined to be $\BR (\BsToKzKzb) = (19.6^{+5.8}_{-5.1} \pm 1.0 \pm 2.0) \times 10^{-6}$, where the first uncertainty is statistical, the second systematic and the third due to the uncertainty of the total number of produced \Bs--\Bsb pairs. The related decay \BzToKzKzb has a branching fraction of $(1.21 \pm 0.16)\times10^{-6}$\cite{PDG2018,Duh:2012ie,Aubert:2006gm} in the world average.

This paper presents measurements of the branching fraction of \BszToKSKS decays using proton-proton collision data collected by the LHCb experiment at centre-of-mass energies $\sqrt{s} = 7$, $8$, or \SI{13}{TeV}. The \BszToKSKS branching fraction is assumed to be half of the \BszToKzKzb branching fraction, as the \Kz\Kzb final state is \CP even. These \BdorBs branching fractions are determined relative to the \BzToKSphi branching fraction, where the notation $\phi$ is used for the $\phi(1020)$ meson throughout. This normalization mode has a corresponding branching fraction equal to half of $\BR(\BzToKzphi) = \num{7.3 +- 0.7e-6}$~\cite{Lees:2012kxa,Chen:2003jfa}, and is chosen for its similarity to the signal mode. Despite the smaller branching fraction, the yield of the normalization mode is much larger than that of the signal mode, because the near-instantaneous \phiz decay can be reconstructed more efficiently than a long-lived \KS, and because for LHCb the production fraction of \Bz mesons is approximately four times that of \Bs mesons. \cite{fsfd,LHCb-PAPER-2018-050}. Throughout this paper, the decays \BszToKSKS and \BzToKSphi are reconstructed using the decays \KSTopippim and \phiToKK.

The paper is structured as follows.
A brief description of the \lhcb detector as well as the simulation and reconstruction software is given in~\cref{sec:detector}.
Signal selection and strategies to suppress background contributions are outlined in~\cref{sec:selection}.
The models to describe the invariant-mass components, the fitting and the normalization procedure are introduced in~\cref{sec:fitstrategy}.
Systematic uncertainties are discussed in~\cref{sec:systematicuncertainties}.
Finally, the results are summarized in~\cref{sec:conclusion}.

\section{LHCb detector}
\label{sec:detector}

The \lhcb detector~\cite{Alves:2008zz,LHCb-DP-2014-002} is a single-arm forward
spectrometer covering the \mbox{pseudorapidity} range $2<\eta <5$,
designed for the study of particles containing \bquark or \cquark
quarks. The detector includes a high-precision tracking system
consisting of a silicon-strip vertex detector (\velo) surrounding the $pp$
interaction region~\cite{LHCb-DP-2014-001}, a large-area silicon-strip detector located
upstream of a dipole magnet with a bending power of about
$4{\mathrm{\,Tm}}$, and three stations of silicon-strip detectors and straw
drift tubes~\cite{LHCb-DP-2013-003,LHCb-DP-2017-001} placed downstream of the magnet.
The tracking system provides a measurement of momentum, \ptot, of charged particles with
a relative uncertainty that varies from 0.5\% at low momentum to 1.0\% at 200\gevc.
The minimum distance of a track to a primary vertex (PV), the impact parameter (IP), 
is measured with a resolution of $(15+29/\pt)\mum$,
where \pt is the component of the momentum transverse to the beam, in\,\gevc.
Different types of charged hadrons are distinguished using information
from two ring-imaging Cherenkov detectors~\cite{LHCb-DP-2012-003}. 
Photons, electrons and hadrons are identified by a calorimeter system consisting of
scintillating-pad and preshower detectors, an electromagnetic
calorimeter and a hadronic calorimeter. Muons are identified by a
system composed of alternating layers of iron and multiwire
proportional chambers.

The online event selection is performed by a trigger~\cite{LHCb-DP-2012-004}, 
which consists of a hardware stage, based on information from the calorimeter and muon
systems, followed by a software stage, which applies a full event
reconstruction. 
At the hardware trigger stage, events are required to contain a muon with high $\pt$ or a hadron, photon or electron with 
high transverse energy in the calorimeters.
In the software trigger, events are selected by a topological \bquark-hadron trigger. At least one charged particle must have a large transverse momentum and be inconsistent with originating from any PV. A two- or three-track
secondary vertex is constructed, which must have a large sum of the $\pt$ of the charged particles and a significant displacement from any PV. 
A multivariate algorithm~\cite{BBDT} is used for
the identification of secondary vertices consistent with the decay
of a \bquark-hadron. This is used to collect both \BszToKSKS and \BzToKSphi decays. In addition to this topological trigger and algorithm, some \BzToKSphi decays are also collected using dedicated $\phi$ trigger requirements that exploit the topology of the \phiToKK decay and apply additional particle identification requirements to the charged kaons.

Simulation is required to model the effects of the detector acceptance and the 
imposed selection requirements. In simulation, $pp$ collisions are generated using
\pythia~\cite{Sjostrand:2006za,*Sjostrand:2007gs} with a specific \lhcb
configuration~\cite{LHCb-PROC-2010-056}.  Decays of hadronic particles
are described by \evtgen~\cite{Lange:2001uf}, in which final-state
radiation is generated using \photos~\cite{Golonka:2005pn}. The
interaction of the generated particles with the detector, and its response,
are implemented using the \geant
toolkit~\cite{Allison:2006ve, *Agostinelli:2002hh} as described in
Ref.~\cite{LHCb-PROC-2011-006}.

\section{Event selection}
\label{sec:selection}

The decays \BszToKSKS and \BzToKSphi are reconstructed using the decay modes \KSTopippim and \phiToKK.\footnote{The inclusion of charge-conjugate processes is implied throughout the paper} The long-lived \KS mesons are
reconstructed in two different categories, depending on whether the \KS meson decays early enough that the pions can be tracked inside the VELO, or whether the \KS meson decays later and its products can only be tracked downstream. These are referred to as \textit{long} and
\textit{downstream} track categories, and are abbreviated as L and D, respectively.
The \KS mesons reconstructed in the long track category have better mass, momentum and vertex resolution than the downstream track category.
However, due to the boost of the \B meson, the lifetime of the \KS meson, and the geometry of the detector, there are approximately twice as many \KS candidates reconstructed in the downstream category than in the long category, before any selections are applied. 

This analysis is based on $\proton\proton$ collision data collected by the \lhcb experiment. Data collected 
in 2011 (2012) were recorded at a center-of-mass energy of $\SI{7}{\tev}$ ($\SI{8}{\tev}$), while in 2015 and 2016 
the center-of-mass energy was increased to $\SI{13}{TeV}$.  Data recorded at center-of-mass energies of 7 and
\SI{8}{\tev} (Run~1) are combined and then treated separately from data recorded at $\SI{13}{TeV}$ (Run~2). Due to low trigger efficiency for $\Bs$ mesons decaying into two downstream \KS mesons, these are discarded from the analysis. Consequently, there are four data categories that are considered in the following---Run~1 LL, Run~1 LD, Run~2 LL and Run~2 LD---and measurements are performed separately in each of these data categories before being combined in the final fit.

Signal \Bs or \Bz candidates are built in successive steps, with individual \KS candidates reconstructed first and then combined. The \KS candidates are constructed by combining two oppositely charged pions that meet certain
requirements on the minimum total momentum and transverse momentum; on 
the minimum \chisqip of the \KS candidate with respect to the associated PV (where \chisqip is defined as the
difference in the impact parameter $\chisq$ of a given PV reconstructed with and without the considered particle);
on the maximum distance of closest approach (DOCA) between the two particles; and on the quality of the vertex fit. An event can have more than one PV, in which case the associated PV is defined as that with which the B candidate forms the smallest value of \chisqip.
The invariant mass of \KS candidates constructed from long (downstream) tracks must be within 35\mevcc (64\mevcc ) of the known \KS mass~\cite{PDG2018}. 
The DOCA between
the two \KS candidates is required to be smaller than \SI{1}{\milli\metre} for the LL category and \SI{4}{\milli\metre} for the LD category. 
Signal \Bs or \Bz candidates are then formed by combining two \KS candidates that result in an
invariant mass close to the known \BdorBs masses, discussed further below and in Sec. \ref{sec:fitstrategy}.

The normalization decay \BzToKSphi is constructed in a similar way. The \phiz meson is constructed by combining two oppositely charged kaon candidates 
that result in an invariant mass within \SI{50}{\mevcc} of the nominal \phiz mass, as a first loose selection. Due to the vanishing lifetime of the \phiz meson, 
the charged kaon candidates are only reconstructed from long tracks, and thus all \phiz are reconstructed in the L category. The \KS meson of the normalization decay can be either L or D, so that the \BzToKSphi decay has both LL and LD reconstructions.

The rest of the candidate selection process consists of a preselection followed by the application of a multivariate classifier, and then some additional selections are applied to further reduce combinatorial background. In the preselection, loose selection requirements are applied to remove specific backgrounds from other $b$-hadron decays and suppress combinatorial background. These backgrounds for the signal and normalization modes are discussed further below. Additional suppression of the combinatorial background is included using a final selection after the multivariate classifier is applied, where particle identification (PID) requirements are added such that all final-state particles must be inconsistent with the muon hypothesis based on the association of hits in the muon stations.

Possible background decays are studied using
simulated samples. For the signal channel, these include: $B_{(s)}^0 \to \KS \pip \pim$; $B_{(s)}^0 \to \KS \pip \Km$
with kaon--pion misidentification; $B_{(s)}^0 \to \KS \Kp \Km$ with double kaon--pion misidentification; and \LbTopKSpi with
proton--pion misidentification. Backgrounds from \decay{\KL}{\pip\pim} decays are negligible.
Applying the \KS mass window requirement to the two-hadron system originating directly from a \bquark-hadron decay reduces the background yields by a factor of
$10$ to $100$, depending on the decay channel. To further suppress the contribution of these modes, a requirement on the distance along the beam axis direction (the $z$-direction) between the
decay vertices of the $\KS$ and $\Bs$ candidates, $\Delta z > \SI{15}{\milli\metre}$, is applied to $\KS$ candidates reconstructed from long tracks for both decay channels.

An additional background comes from the requirements used to identify \KS candidates, which may also select \Lz baryons due to their long flight distance. The $\Lz \to \proton \pim$ decays are excluded by
changing the mass hypothesis of one pion candidate to the proton hypothesis, reconstructing the invariant mass,
$m(\proton \pim)$,  and tightening the pion PID requirement in an \SI{8}{\mevcc} mass window around the known $\Lz$ mass. This procedure is carried out for each pion from each \KS candidate, in both the signal and normalization channels.

For the normalization channel \BzToKSphi, the decays $B_{(s)}^0 \to \KS h^{(')+} h^{(')-}$ with $h^{(')\pm} =
\pi^{\pm}, \kaon^{\pm}$ are suppressed by requiring the invariant mass of the combination of the two final-state kaons to
be close to the \phiz mass. The largest contributions are expected from the decay channel $\Bs \to \KS \pip \Km$ with a
fraction of about $\SI{1}{\percent}$ compared to \BzToKSphi decays. This is reduced to a negligible level by applying
PID requirements to the kaon candidates.
The partially reconstructed decays $\Bz \to \phiz \Kstarz$ and $\Bp \to \phiz \Kstarp$, with $\Kstarz \to \KS \piz$ and $\Kstarp \to
\KS \pip$, share the same decay topology as the normalization channel when omitting the pion that originates from
decay of the $K^{*}$ resonance and have a higher branching fraction than the normalization decay. Due to the missing
particle, the $B$ candidates have a kinematic upper limit on their masses of about \SI{5140}{\mevcc}. Therefore, the mass
window to determine the yield of the normalization channel is set to $\SI{5150} <~m(\KS \Kp \Km) <
\SI{5600}{\mevcc}$ to fully exclude these contributions. 

Further separation of signal from combinatorial background is achieved using the XGBoost implementation~\cite{XGBOOST} of the Boosted Decision Tree (BDT) algorithm~\cite{Breiman}. For the training,
simulated signal (normalization) decays are used as signal proxy, while the upper mass sideband
$m(\KS\KS) > \SI{5600}{\mevcc}$ ($m(\phiz\KS) > \SI{5600}{\mevcc}$) in data is utilized as background proxy. 
To account for differences in data and simulation, the simulated decays are weighted 
in the $B$ meson production kinematics and detector occupancy (represented by the number of tracks in the event)
to match data distributions. 

The BDT exploits the following observables:
the flight distance, IP and $\chi^2_{\text{IP}}$ of the $B$ and \KS candidates with respect to all primary vertices, as well as 
the decay time, the momentum, transverse momentum and pseudorapidity of the $B$ candidate.
This set of quantities is chosen such that they have a high separation power between signal and background and are not
directly correlated to the invariant mass. 
The same procedure is applied to the \BzToKSphi data samples. 

In order to choose the optimal threshold on the BDT response, the figure of merit
$\varepsilon_{\text{sig}}/(3/2 + \sqrt{N_{\text{bkg}}})$~\cite{Punzi:2003bu} is used for the signal mode, where the value $3/2$ corresponds to a target 3 sigma significance and $\varepsilon_{\text{sig}}$ is the signal efficiency of the selection, determined from simulation. The figure of merit $N_{\text{sig}}/\sqrt{N_{\text{sig}} + N_{\text{bkg}}}$ is used for the normalization channel to minimize the uncertainty on the yield.
So as not to bias the determination of the signal yield, the candidates in the signal region were 
not inspected until the selection was finalized. Consequently, the expected background yield $N_{\text{bkg}}$
is calculated by interpolating the result of an exponential fit to data sidebands,
$\SI{5000} <~m(\KS\KS) < \SI{5230}{\mevcc}$, and $\SI{5420} <~m(\KS\KS) < \SI{5600}{\mevcc}$ into the signal region.
For the normalization channel, the variation of the expected signal yield $N_{\text{sig}}$ as a function of the BDT response threshold is determined from simulation, while the absolute normalization is set from a single fit to the data.

The figure of merit optimization is performed simultaneously with respect to the BDT classifier output and an observable based on PID information for long track candidates, where the latter observable is corrected using a resampling from data calibration samples~\cite{LHCb-DP-2018-001} to minimize differences in data and simulation. 
As a last selection step, the invariant-mass windows of $m(\pip\pim)$ and $m(\Kp \Km)$ are tightened to further suppress combinatorial background. 
Finally, multiple candidates, which occur in about 1 in 10000 of all events, are removed randomly so that each event contains only one signal candidate.

\section{Fit strategy and results}
\label{sec:fitstrategy}

\begin{figure}[tb]
	\centering
	\includegraphics[width=0.49\linewidth]{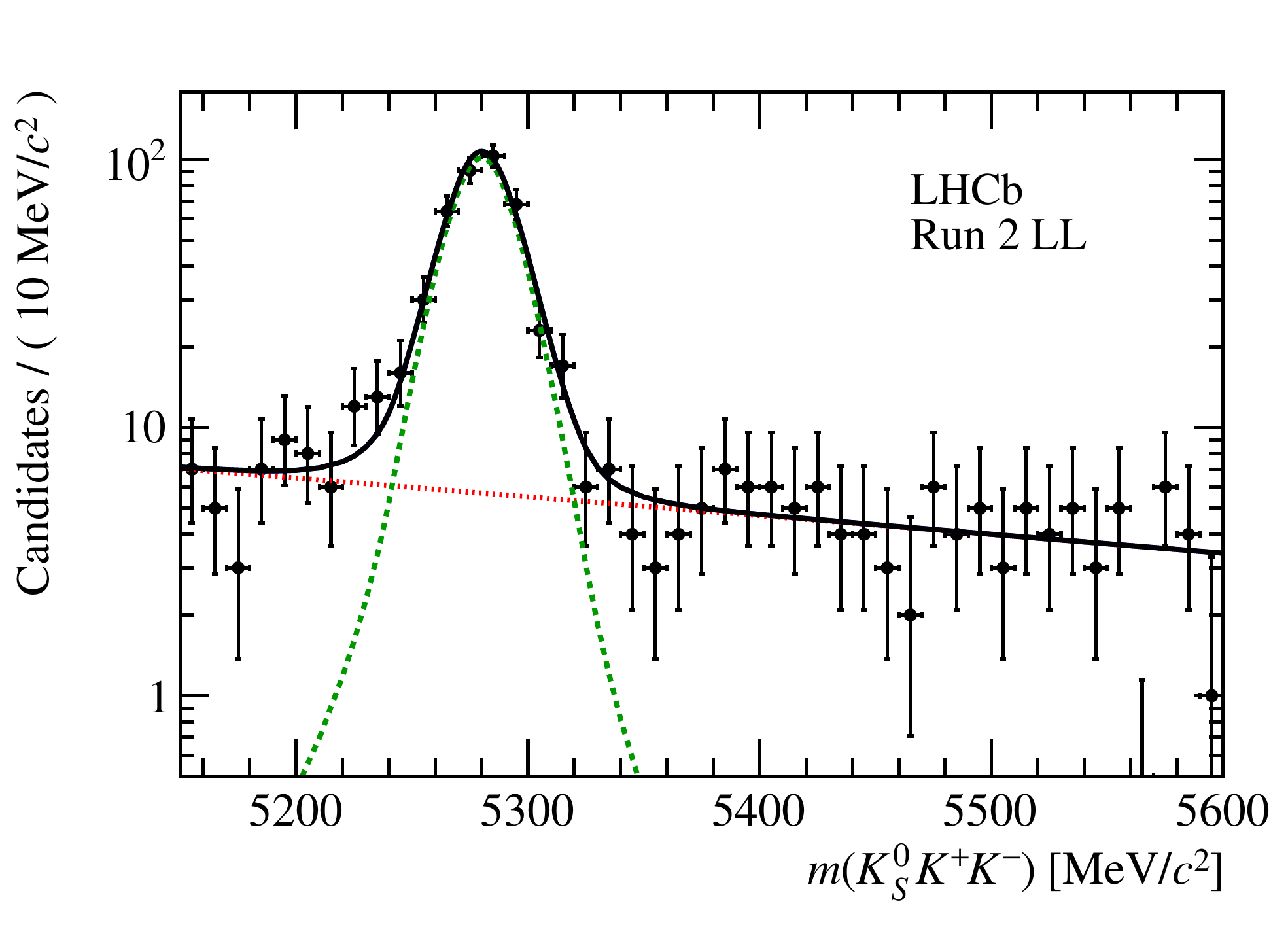}
	\includegraphics[width=0.49\linewidth]{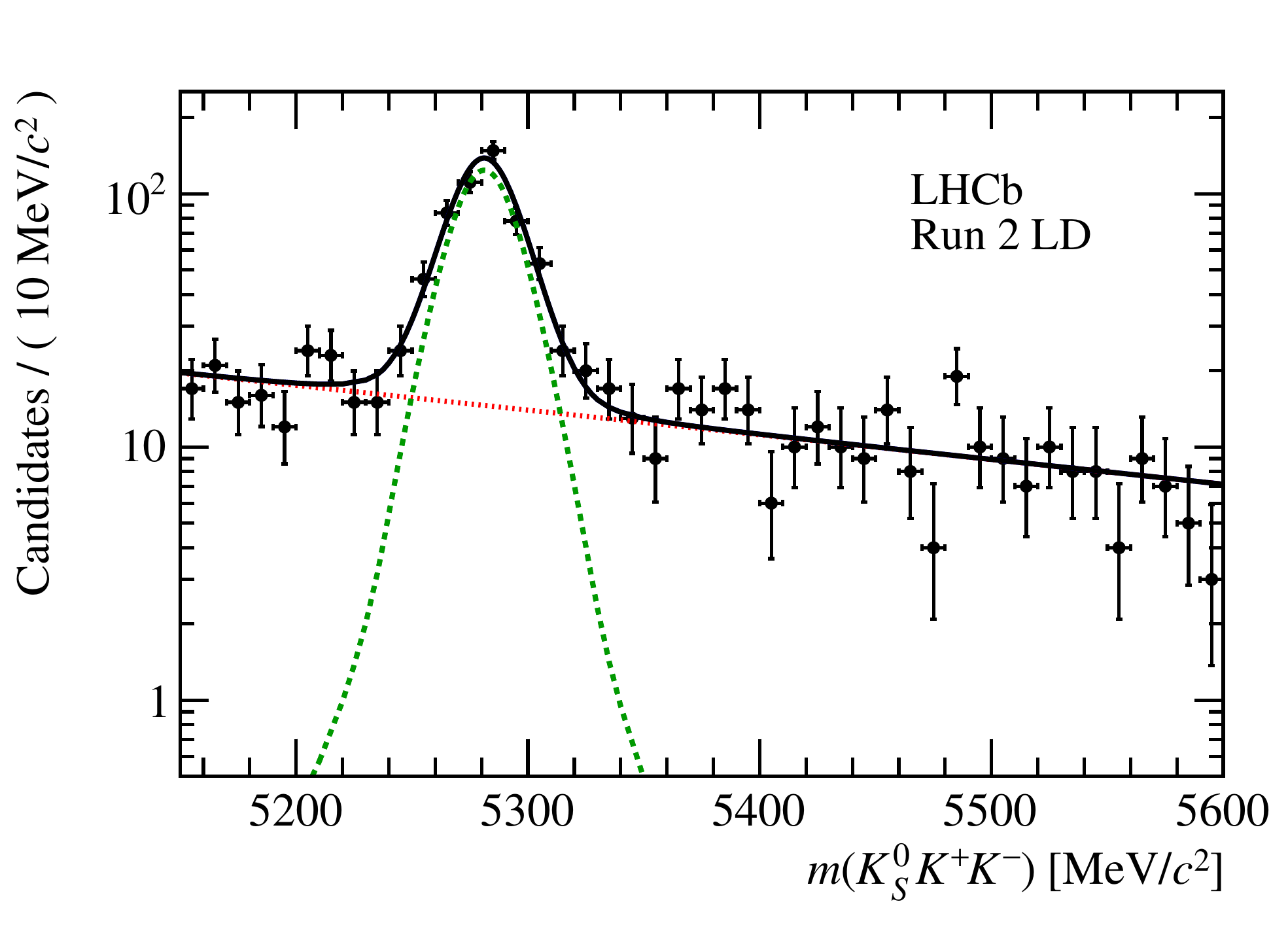}
	\caption{\small Fits to the invariant-mass distribution $m(\KS\Kp\Km)$ of the normalization decay channel. The black curve represents the complete model, the \BzToKSphi component is given in green
		(dashed), while the background component in shown in red (dotted).
	}
	\label{fig:Run2KKKS}
\end{figure}

For the normalization channel, the total \BzToKSphi yield is obtained from extended unbinned maximum
likelihood fits to the reconstructed $B^0$ mass in the range \SIrange{5150}{5600}{\mevcc}, separately for each data
sample and reconstruction category. The signal component is modelled by a Hypatia function with power-law tails on both sides~\cite{Santos:2013gra}, where the tail
parameters are fixed to values obtained from fits to simulated samples. The mean, width and signal yield parameters are free to vary in the fit. An exponential function with a free slope parameter models the combinatorial background. 
To account for non-\phiz contributions to the \BzToKSphi yield, 
a subsequent fit is performed to the $m(\Kp\Km)$ distribution, which is background-subtracted using the \splot technique~\cite{Pivk:2004ty} and where the $m(\Kp\Km\KS)$ distribution is used as the discriminating variable.
The signal \phiz component of the $m(\Kp\Km)$ fit is modelled by a relativistic Breit--Wigner function~\cite{Kycia:2018hyf} convolved with a Gaussian function to take into
account the resolution of the detector, while the non-\phiz contributions are described by an exponential function. The
slope parameter of the latter model is Gaussian-constrained to the results obtained from fits to the simulation of
$f_0(980) \to \Kp \Km$ decays, which is found to better describe the observed distribution than a phase-space model. The measured yields for the normalization channel are shown in the last row of Table \ref{tab:fitresults_bs2ksks}. Plots of the $m(\Kp\Km\KS)$ distributions for the Run 2 LL and LD samples are shown in Figure \ref{fig:Run2KKKS}. The remaining $m(\Kp\Km\KS)$ distributions and the $m(\Kp\Km)$ distributions are shown in the \hyperref[appendix]{Appendix}.

A Hypatia function is used to model the $m(\KS\KS)$ distribution of signal \BsToKSKS decays. All shape parameters
are fixed to values obtained from fits to simulated samples. To account for resolution differences between simulation
and data, the width is scaled by a factor---determined from the normalization channel---which takes values in the range 1.05 to 1.20 depending on the data sample. To model the \BzToKSKS signal component, the same signal shape
is duplicated and shifted by the $\Bs-\Bz$ mass difference~\cite{LHCb-PAPER-2015-010}. The
background component is modelled by an exponential function with a free slope parameter. 

In contrast to the normalization channel, where each data category is fitted individually, a
simultaneous fit to the $m(\KS\KS)$ distribution of the four data categories (Run 1 LL, Run 1 LD, Run 2 LL, Run 2 LD) is performed in the range \SIrange{5000}{5600}{\mevcc}. Two
parameters are shared across all categories in the simultaneous fit, the ratio of the \BzToKSKS and \BsToKSKS yields $f_{\Bz/\Bs}$ and the
branching fraction $\BR(\BsToKSKS)$, which is itself related to the signal yield of each data category 
via the relation
\begin{align}
\label{eq:BRnormalization}
\BR(\BsToKSKS) &=  
\frac{\varepsilon_{\phi\KS,\,i}}{\varepsilon_{\KS\KS,\,i}} 
\frac{\fracd}{\fracs} 
\frac{\BR(\phiToKK)}{\BR(\KSTopippim)}
\frac{\BR(\BzToKSphi)}{N_{i}(\BzToKSphi)}
\cdot N_{i}(\BsToKSKS) \\
&\equiv\alpha_{i} \cdot N_{i}(\BsToKSKS) \nonumber \,,
\end{align}
where the normalization constant $\alpha_{i}$ is introduced for each data category sample $i$. While the selection efficiencies $\varepsilon$ and signal yields $N$ are determined in the present
analysis, external sources are used for the ratio of fragmentation fractions $\sfrac{\fracd}{\fracs}$~\cite{fsfd,LHCb-PAPER-2018-050}, and the branching fractions
$\BR(\BzToKSphi)$,  $\BR(\KSTopippim)$ and $\BR(\phiToKK)$~\cite{PDG2018}. To increase the robustness of the fit, the $\alpha_{i}$ constants are Gaussian constrained within their uncertainties, excluding the uncertainties from the external constants. These external uncertainties are instead applied directly to the final branching ratio measurement.

The efficiency ratio, $\varepsilon_{\phi\KS}/\varepsilon_{\KS\KS}$, is determined from simulation and corrected using data control samples. This ratio is found to be approximately equal to 30 in all data samples except the Run~1 LD sample, where it is twice as large  due to lower trigger efficiency for downstream tracks in this sample.

\begin{table}[b]
	\centering
	\caption{Results of the simultaneous fit to the invariant mass of the $\KS\KS$ system. The fit results for \BR and $f_{\Bz/\Bs}$ are shared among all data categories. The given uncertainties are statistical only. The normalization constant $\alpha$ and the corresponding normalization channel yields $N_{\text{norm}}$ are shown for reference.}
	\label{tab:fitresults_bs2ksks}
	\resizebox{\textwidth}{!}{
		\begin{tabular}{lS[table-format=2.2(2)]S[table-format=2.2(2)]S[table-format=2.2(2)]S[table-format=2.2(2)]l}
			\toprule
			{} &  {Run 1 LL} & {Run 1 LD} &  {Run 2 LL} &  {Run 2 LD} &           {Status} \\
			{Parameter}        &              &             &              &      &                   \\
			\midrule
			{\BR} $(\times 10^{-6})$              &    \multicolumn{4}{c}{8.3 $\pm$ 1.6}    &            Free \\
			{$f_{\Bz/\Bs}$}    &  \multicolumn{4}{c}{0.30 $\pm$ 0.13}    &                   Free \\
			{$N_{\Bs}$} &    4.3 +- 1.0 &   2.1 +- 0.5 &   12.8 +- 2.7 &   12.4 +- 2.7 &    $\BR / \alpha$ \\
			{$N_{\Bz}$} &    	1.3 +- 0.5 &  	 0.63 +- 0.26 &  	 3.8 +- 1.5 & 	 3.7 +- 1.5 &    	$f_{\Bz/\Bs} \times \BR / \alpha$ \\
			{$N_{\text{bkg}}$} &   10.4 +- 3.5 &   3.5 +- 2.2 &    7.2 +- 3.0 &       13 +- 4 &            Free \\
			{$\alpha$}  $(\times 10^{-6})$       &  1.90 +- 0.21 &   3.9 +- 0.5 &  0.65 +- 0.05 &  0.66 +- 0.05  &  Gaussian constr. \\
			{$N_{\text{norm}}$} & \multicolumn{1}{c}{\tablenum[table-format=3(2)]{179 +- 18}} & \multicolumn{1}{c}{\tablenum[table-format=3(2)]{178 +- 22}} & \multicolumn{1}{c}{\tablenum[table-format=3(2)]{316 +- 25}} & \multicolumn{1}{c}{\tablenum[table-format=3(2)]{400 +- 31}} & Included in $\alpha$ \\
			\bottomrule
		\end{tabular}
	}
\end{table}

\begin{figure}[tb]
	\centering
	\includegraphics[width=0.69\linewidth]{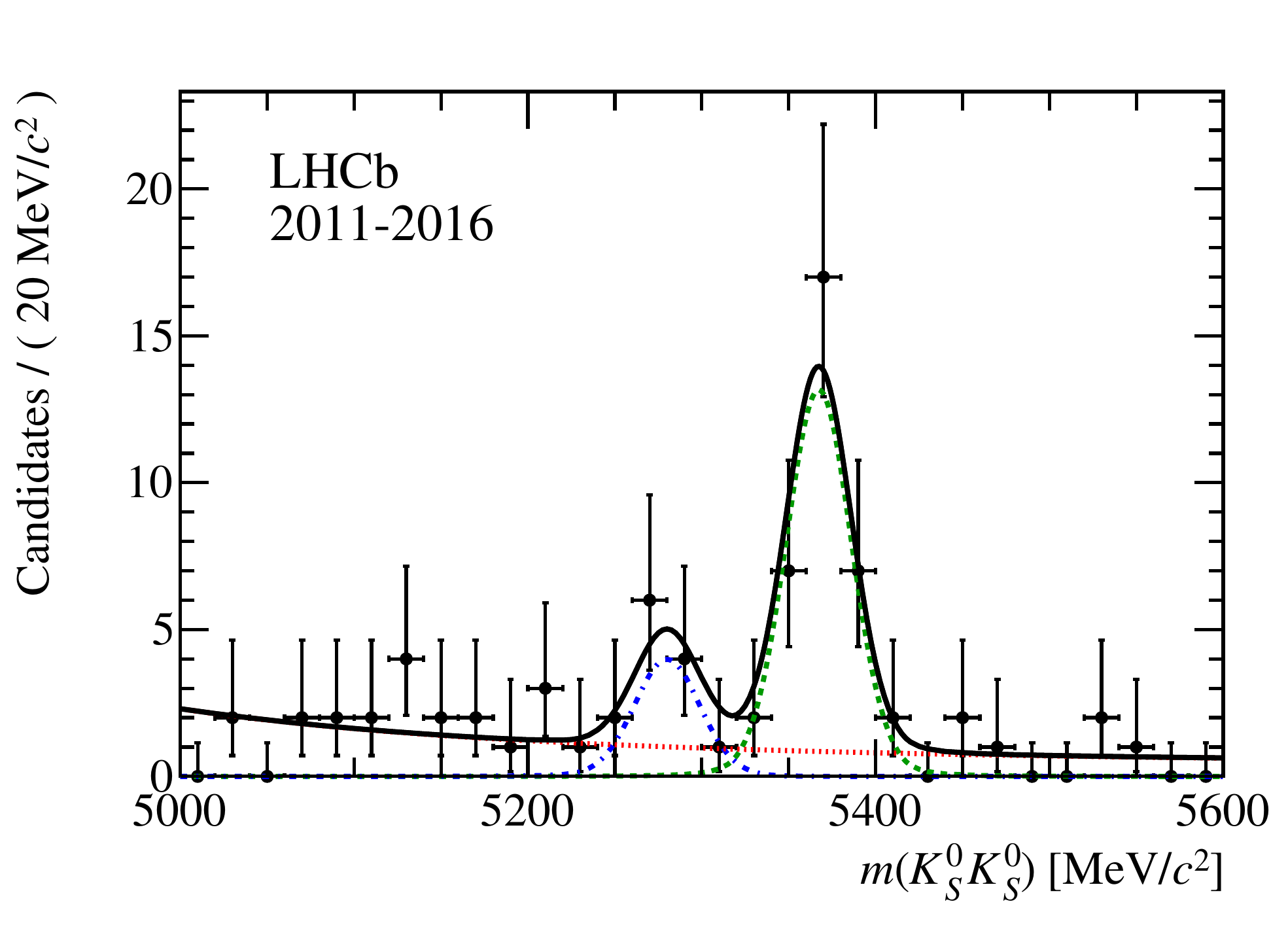}
	\caption{
		\small
		Combined invariant-mass distribution $m(\KS \KS)$ of the signal decay channel. The black (solid) curve represents
		the complete model, the \Bs signal component is given in green (dashed), the smaller \Bz signal is given in blue
		(dash-dotted) and the background component in red (dotted).
	}
	\label{fig:massfit_bs2ksks}
\end{figure}

The fit results are shown in~\cref{fig:massfit_bs2ksks}. The results of the simultaneous mass fit are given in~\cref{tab:fitresults_bs2ksks}, 
yielding a branching fraction of
$\BR(\BsToKSKS) = \num{8.3 +- 1.6 e-6}$, where the uncertainty is statistical only. The $\BsToKSKS$ yield is around 32. The ratio of the branching fractions of the signal and normalization modes $\BR(\BsToKSKS)/\BR(\BzToKSphi)$ can also be calculated by removing the contribution of the world-average value of $\BR(\BzToKSphi)$ from the fit result. This yields a combined branching fraction ratio $\BR(\BsToKSKS)/\BR(\BzToKSphi) = \num{2.3 +- 0.4}$, where the uncertainty is statistical only.

From the same fit, the relative fraction of \BzToKSKS decays, $f_{\Bz/\Bs}= \num{0.3 +- 0.13}$ is also determined. Given that the final-state particles and selections applied to the \KS candidates are the same for both modes, the ratio of selection efficiencies is equal to one, so that $f_{\Bz/\Bs}$ can be converted to a ratio of branching fractions by multiplying by $\sfrac{f_s}{f_d}$. The calculated value of $\BR(\BzToKSKS)/\BR(\BsToKSKS)$ is $\num{7.5 +- 3.1 e-2}$, where the uncertainty is statistical only.

The significances of the \BsToKSKS and \BzToKSKS signal yields are estimated relative to a background-only hypothesis using Wilks' theorem~\cite{Wilks:1938dza}. The observed signal yield of 32 \BsToKSKS decays has a large significance of $\SI{8.6}{\sigma}$ ($\SI{6.5}{\sigma}$ including the effect of systematic uncertainties), while the smaller \BzToKSKS signal yield has a significance of $\SI{3.5}{\sigma}$ including systematic uncertainties.

\section{Systematic uncertainties}
\label{sec:systematicuncertainties}

Each source of systematic uncertainty is evaluated independently and expressed as a relative uncertainty on the
branching fraction of \BsToKSKS decays. A complete list is given in~\cref{tab:syst_uncertainties}. The uncertainties are grouped into three general categories: fit and weighting uncertainties, PID uncertainties, and detector and trigger uncertainties.

\begin{table}[tb]
	\centering
	\caption{
		\small
		All systematic uncertainties on the \BsToKSKS branching fraction, presented as relative measurements. The last row shows the combined systematic uncertainty for each data sample.
	}
	\label{tab:syst_uncertainties}
	\begin{tabular}{lcccc}
		\toprule
		{} &  {Run 1, LL} &  {Run 1, LD} &  {Run 2, LL} &  {Run 2, LD} \\
		{Systematic uncert.}            &              &              &              &              \\
		\midrule
		{Fit bias}           &        0.059 &        0.059 &        0.059 &        0.059 \\
		{Fit model choice}   &        0.022 &        0.033 &        0.015 &        0.013 \\
		{Fit model parameters} &        0.026 &        0.026 &        0.026 &        0.026 \\
		{BDT}                &        0.023 &        0.040 &        0.014 &        0.031 \\
		{PID}                &        0.007 &        0.008 &        0.026 &        0.026 \\
		{Hardware trigger}   &        0.063 &        0.062 &        0.063 &        0.062 \\
		{Software trigger}   &        0.065 &        0.106 &        0.008 &        0.026 \\
		{Trigger misconfig.}  &        ---   &        ---   &        0.007 &        0.004 \\
		{\pipm/\Kpm hadronic interaction}     &        0.005 &        0.005 &        0.005 &        0.005 \\
		{VELO misalignment}               &        0.008 &        0.008 &        0.008 &        0.008 \\
		\midrule
		{Total}              &        0.116 &        0.149 &        0.097 &        0.103 \\
		\bottomrule
	\end{tabular}
\end{table}

Multiple different fit uncertainties are considered. Uncertainty from possible bias in the combined fit to all four data samples can be estimated using pseudoexperiments generated and fitted according to the default fit model. In each pseudoexperiment, the number of signal candidates is drawn from a Poisson distribution with a mean
determined from the baseline fit result. A relative average difference between the generated and fitted branching fraction of \SI{5.9}{\percent}
is determined and conservatively assigned as a systematic uncertainty. The same procedure is performed for the \BzToKSKS
component, yielding a possible bias of \SI{1.6}{\percent}. To ensure a conservative approach, the \SI{5.9}{\percent} value from \BsToKSKS is also applied as the systematic for \BzToKSKS.

Another systematic uncertainty in the fitting process arises from the specific fit model choice, which
is quantified by the use of alternative probability density functions to describe the invariant-mass distributions. The reconstructed-mass shapes for \Bz and \Bs
mesons are modelled by the sum of two Crystal Ball functions~\cite{Skwarnicki:1986xj}. For
the fit to the $m(\Kp\Km)$ distribution, the \phiz meson is modelled by a non-relativistic Breit--Wigner function.
For the normalization channel, the relative yield difference when refitting the data is taken as the systematic uncertainty, 
while for the \BsToKSKS decay pseudoexperiments are used to estimate the impact of mismodeling the shape of the signal component. 
The systematic uncertainty due to the choice of fit model is then the sum in quadrature 
of these variations, yielding values of \SI{1.3}{\percent} to \SI{3.3}{\percent} depending on the data category. 
Another systematic uncertainty of \SI{2.6}{\percent}, evaluated with a similar procedure, is assigned due to fixing certain shape parameters to values obtained in fits to simulated samples. 

Additionally, not all differences between data and simulation can be accounted for using weights in the BDT training. As a conservative upper
limit of this effect, the signal efficiency is calculated with and without weights, and the differences between these efficiencies are treated as
a systematic uncertainty. 
This systematic uncertainty is larger by a factor of about 2 for data categories containing a downstream \KS candidate than in those that contain only long \KS candidates,
indicating a stronger dependence of the LD channel efficiencies on the weighting. 

Three sources of systematic uncertainties from PID efficiencies are considered. The effect of the finite size of the signal 
simulation samples is evaluated using the bootstrap method~\cite{Efron:526679} for each simulation category and
calculating the variance of the signal efficiency. A second systematic uncertainty is calculated by varying the model used to resample PID calibration data, 
and the relative difference in the signal efficiency is taken as a systematic uncertainty, though this effect is small compared to the previous source. Finally, the 
flight distance of the \KS candidate is not considered in the resampling process, 
while the PID efficiency does exhibit some correlation with this variable. A systematic uncertainty is
calculated by reweighting the PID distributions in bins of the \KS flight distance, and calculating the relative signal efficiency on resampled simulation and resampled and reweighted simulation.  
The combined PID systematic uncertainty is given by summing over the three effects in quadrature, 
which is below \SI{1}{\percent} for the Run~1 samples and below \SI{3}{\percent} for the Run~2 samples. 

Systematic uncertainties in the trigger system are divided into hardware and software trigger uncertainties. For the hardware trigger stage, the efficiency taken from simulation is compared with data calibration samples. The calibration data is used to correct the simulated efficiencies, and the resulting \SI{6}{\percent} relative difference in efficiency between the signal and normalization modes is treated as a systematic uncertainty.
For the inclusive $B$ software trigger, possible differences in efficiency
between the signal and normalization channels are obtained by reweighting the \BzToKSphi simulation to match the \BsToKSKS simulation and
calculating the relative efficiency difference between the raw and reweighted distributions, yielding a systematic
uncertainty of about \SI{2}{\percent}. An additional, larger systematic uncertainty is also included to account for the dedicated \phiz trigger requirements, which are
only used for the normalization channel. Again, weighted \BzToKSphi data are used to evaluate a relative efficiency
difference between simulation and data, multiplied by the fraction of events solely triggered by the dedicated \phiz trigger requirements. 
The systematic uncertainty is about \SIrange{5}{10}{\percent} in Run 1, but about 5 times smaller for
Run 2. This is because the topological $b$-hadron trigger is more efficient in Run~2 so that there are far fewer events triggered only by the  dedicated \phiz trigger. An additional systematic uncertainty less than $\SI{1}{\percent}$ is assigned to account for a small known misconfiguration of the
trigger during Run 2 data taking. 

Two additional detector-related uncertainties are considered. A relative uncertainty of \SI{0.5}{\percent} is assigned due to
the different hadronic interaction probabilities between pions and kaons in data and simulation, and a relative uncertainty of \SI{0.8}{\percent} is also introduced to account for a
possible misalignment in the downstream positions of the vertex detector.

The combined systematic uncertainty is determined by using a weighted average of the total systematic uncertainty for each data category, where the weighting is based on the \Bs signal yield for each category, obtained from the nominal combined fit for the branching fraction. This value is then combined with the systematic uncertainties due to the \phiToKK and \KSTopippim branching fractions, to produce an overall systematic uncertainty of \SI{10.7}{\percent}. The systematic uncertainties due to $\BR(\BzToKSphi)$ or $\sfrac{f_s}{f_d}$ are provided separately when necessary. The total systematic uncertainty in the measurement of the \Bz branching fraction is also \SI{10.7}{\percent}.

These measurements of the branching ratio are calculated using the time-integrated event yield, without taking into account \Bs--\Bsb mixing effects. The conversion into a branching ratio that is independent of \Bs--\Bsb mixing can be performed according to the computation given in Ref.~\cite{PhysRevD.86.014027}, where $\mathcal{A}^f_{\Delta \Gamma}$ is calculated from the decay amplitudes of the $B_s^\text{H}$ and $B_s^\text{L}$ states. In this work, the simulation is generated using the average \Bs lifetime, corresponding to the $\mathcal{A}^f_{\Delta \Gamma}=0$ scenario. For this scenario the mixing-corrected SM prediction of the branching ratio is equivalent to the quoted time-integrated branching ratio within uncertainties, because the impact of the scaling from $\Delta\Gamma_s/\Gamma_s=0.135\pm 0.008$~\cite{PDG2018} is small.

Considering that the final state of the decay is \CP-even, the relevant decay lifetime of the \Bs is expected to be closer to that of the $B_s^\text{L}$ state, corresponding to a SM prediction of $\mathcal{A}^f_{\Delta \Gamma}$ close to $-1$. This change in lifetime corresponds to a change in the expected efficiency of the \BsToKSKS reconstruction of approximately \SI{-4.5}{\percent} for $\mathcal{A}^f_{\Delta \Gamma}=-1$, or \SI{+4.5}{\percent} for the less-likely $\mathcal{A}^f_{\Delta \Gamma}=1$.  These scaling factors are not included in the systematic uncertainty for the time-integrated branching ratios presented below.

\section{Conclusion}
\label{sec:conclusion}

Data collected by the LHCb experiment in 2011--2012 and 2015--2016 was used to measure the \BsToKSKS branching fraction. The measured ratio of this branching fraction relative to that of the normalization channel is
\begin{align*}
\frac{\BR(\BsToKSKS)}{\BR(\BzToKSphi)} = 2.3 \pm 0.4 \,(\text{stat}) \pm 0.2 \,(\text{syst}) \pm 0.1
\,(\sfrac{f_s}{f_d}),
\end{align*}
where the first uncertainty is statistical, the second is systematic, and the third is due to the ratio of hadronization fractions. This is compatible with the ratio $\BR(\BsToKSKS)/\BR(\BzToKSphi) = 2.7 \pm 0.9$ calculated from the current world average values~\cite{PDG2018}.

From this measurement, the \BsToKSKS branching fraction is determined to be
\begin{align*}
\BR(\BsToKSKS) = [8.3 \pm 1.6 \,(\text{stat}) \pm 0.9 \,(\text{syst}) \pm 0.8 \,(\text{norm}) \pm 0.3 
\,(\sfrac{f_s}{f_d})] \times 10^{-6} ,
\end{align*}
where the first uncertainty is statistical, the second is systematic, and the third and fourth are due to the normalization channel branching fraction
and the ratio of hadronization fractions $\sfrac{f_s}{f_d}$. This result is the most precise to date and is compatible with SM predictions~\cite{Williamson:2006hb,Beneke:2003zv,PhysRevD.76.074018,PhysRevD.89.074046}
and the previous measurement from the \belle collaboration~\cite{PhysRevLett.116.161801}.

In the same combined fit used for the \BsToKSKS measurement, the fraction of \BzToKSKS decays is also determined. Using this measured fraction of yields, the branching fraction of \BzToKSKS decays measured relative to \BsToKSKS decays is found to be
\begin{align*}
\frac{\BR(\BzToKSKS)}{\BR(\BsToKSKS)} = [7.5 \pm 3.1 \,(\text{stat}) \pm 0.5 \,(\text{syst}) \pm 0.3 \,(\sfrac{f_s}{f_d})] \times 10^{-2},
\end{align*}
where the first uncertainty is statistical, the second is systematic, and the third is due to the ratio of hadronization fractions. For comparison, calculating $\BR(\BzToKSKS)/\BR(\BsToKSKS)$ based on world average-values~\cite{PDG2018} yields $(6.0 \pm 2.0)\%$, which is compatible with the obtained result.

The \BzToKSKS branching fraction relative to the \BzToKSphi normalization mode is determined to be 
\begin{align*}
\frac{\BR(\BzToKSKS)}{\BR(\BzToKSphi)} = 0.17 \pm 0.08 \,(\text{stat}) \pm 0.02 \,(\text{syst}),
\end{align*}
where the first uncertainty is statistical, and the second is systematic.

\section*{Acknowledgements}
\label{sec:acknowledgements}

\noindent We express our gratitude to our colleagues in the CERN
accelerator departments for the excellent performance of the LHC. We
thank the technical and administrative staff at the LHCb
institutes.
We acknowledge support from CERN and from the national agencies:
CAPES, CNPq, FAPERJ and FINEP (Brazil); 
MOST and NSFC (China); 
CNRS/IN2P3 (France); 
BMBF, DFG and MPG (Germany); 
INFN (Italy); 
NWO (Netherlands); 
MNiSW and NCN (Poland); 
MEN/IFA (Romania); 
MSHE (Russia); 
MinECo (Spain); 
SNSF and SER (Switzerland); 
NASU (Ukraine); 
STFC (United Kingdom); 
DOE NP and NSF (USA).
We acknowledge the computing resources that are provided by CERN, IN2P3
(France), KIT and DESY (Germany), INFN (Italy), SURF (Netherlands),
PIC (Spain), GridPP (United Kingdom), RRCKI and Yandex
LLC (Russia), CSCS (Switzerland), IFIN-HH (Romania), CBPF (Brazil),
PL-GRID (Poland) and OSC (USA).
We are indebted to the communities behind the multiple open-source
software packages on which we depend.
Individual groups or members have received support from
AvH Foundation (Germany);
EPLANET, Marie Sk\l{}odowska-Curie Actions and ERC (European Union);
ANR, Labex P2IO and OCEVU, and R\'{e}gion Auvergne-Rh\^{o}ne-Alpes (France);
Key Research Program of Frontier Sciences of CAS, CAS PIFI, and the Thousand Talents Program (China);
RFBR, RSF and Yandex LLC (Russia);
GVA, XuntaGal and GENCAT (Spain);
the Royal Society
and the Leverhulme Trust (United Kingdom).

\appendix
\section*{Appendix: Normalization channel fits}
\label{appendix}
\begin{figure}[H]
	\centering
	\includegraphics[width=0.49\linewidth]{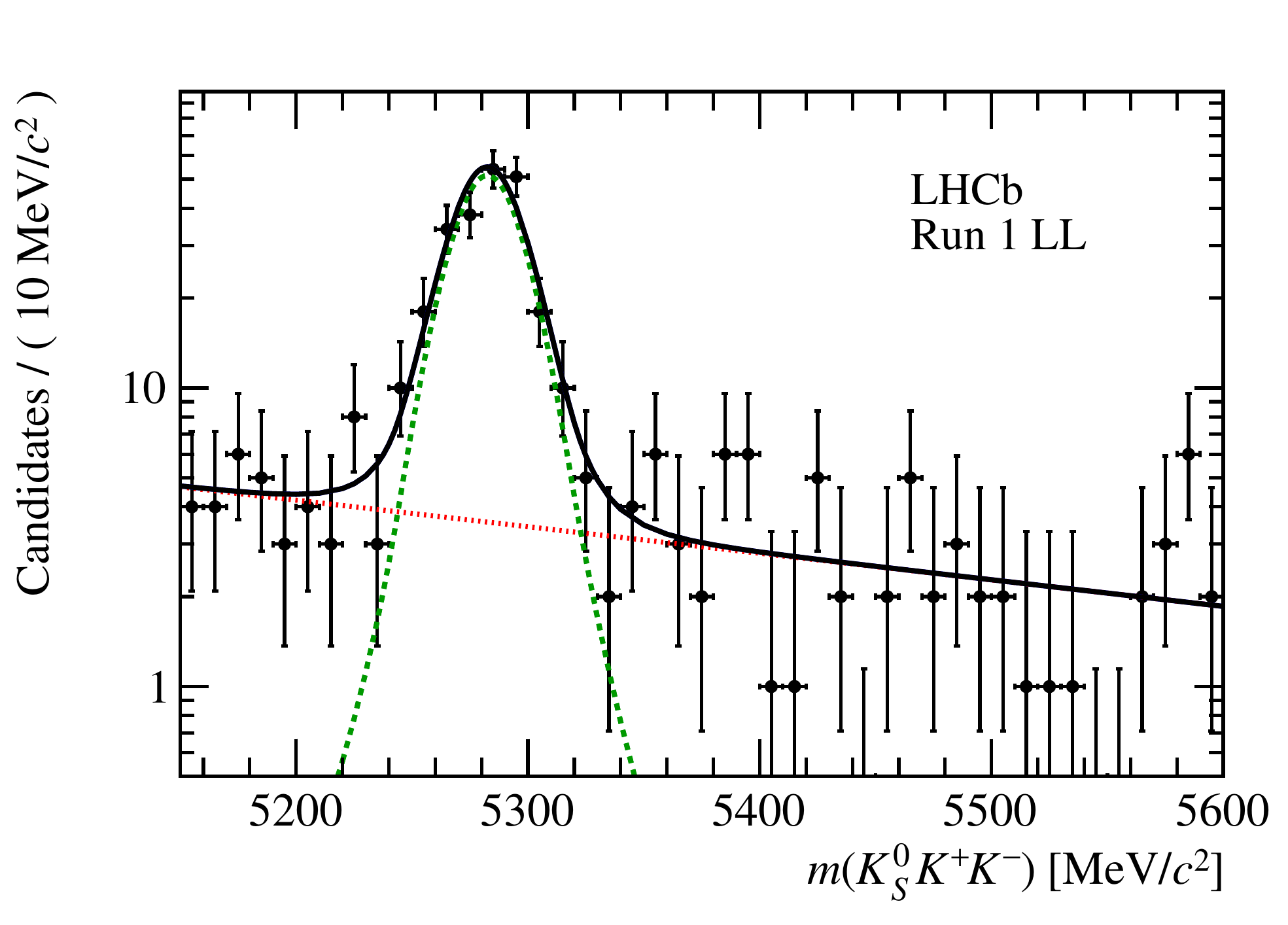}
	\includegraphics[width=0.49\linewidth]{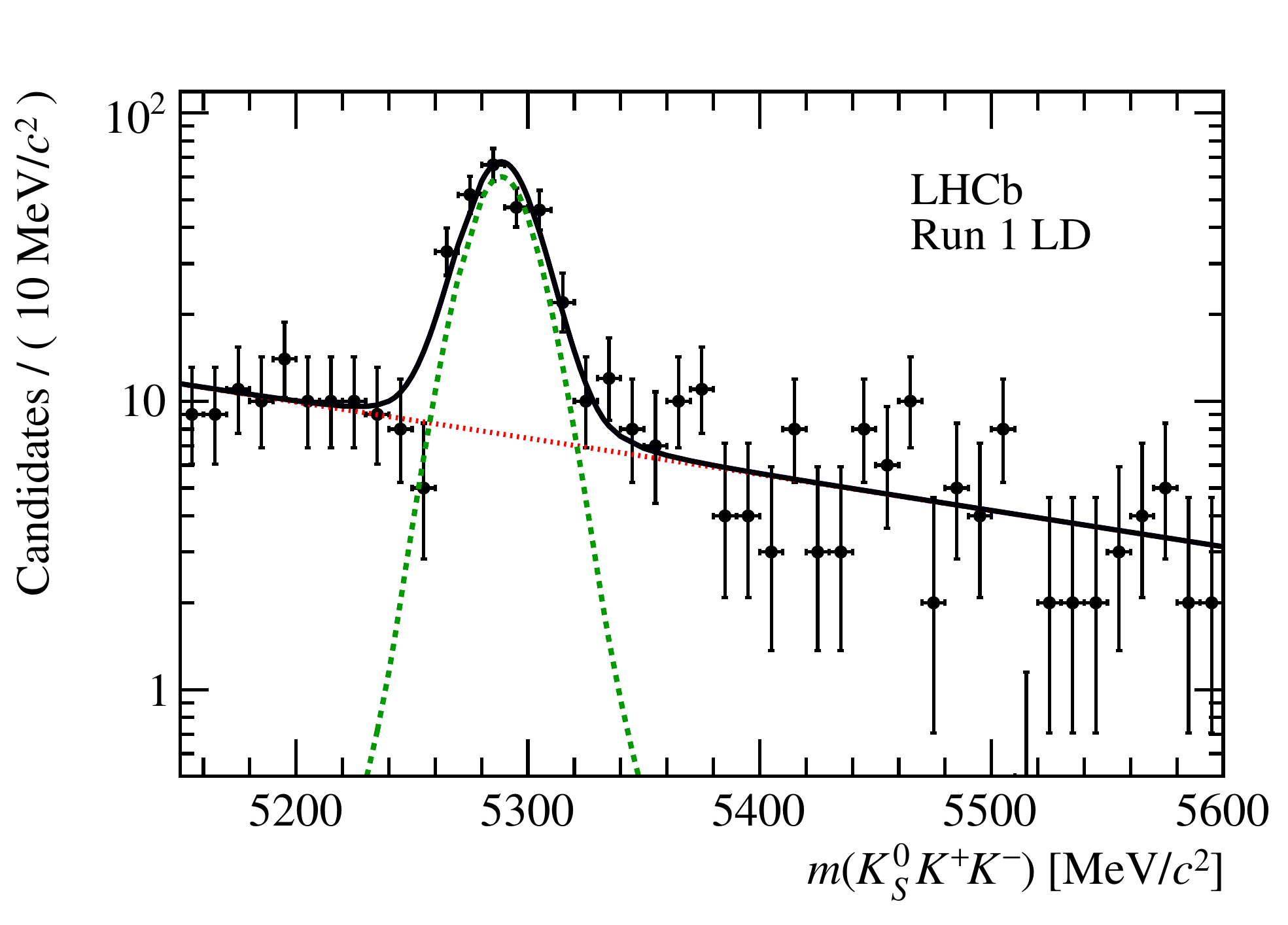}
	\caption{\small Fits to the invariant-mass distribution $m(\KS\Kp\Km)$ of the normalization decay channel. The black curve represents the complete model, the \BzToKSphi component is given in green
		(dashed), while the background component in shown in red (dotted).
	}
	\label{fig:massfit_bd2phiks}
\end{figure}

\begin{figure}[H]
	\centering
	\includegraphics[width=0.49\linewidth]{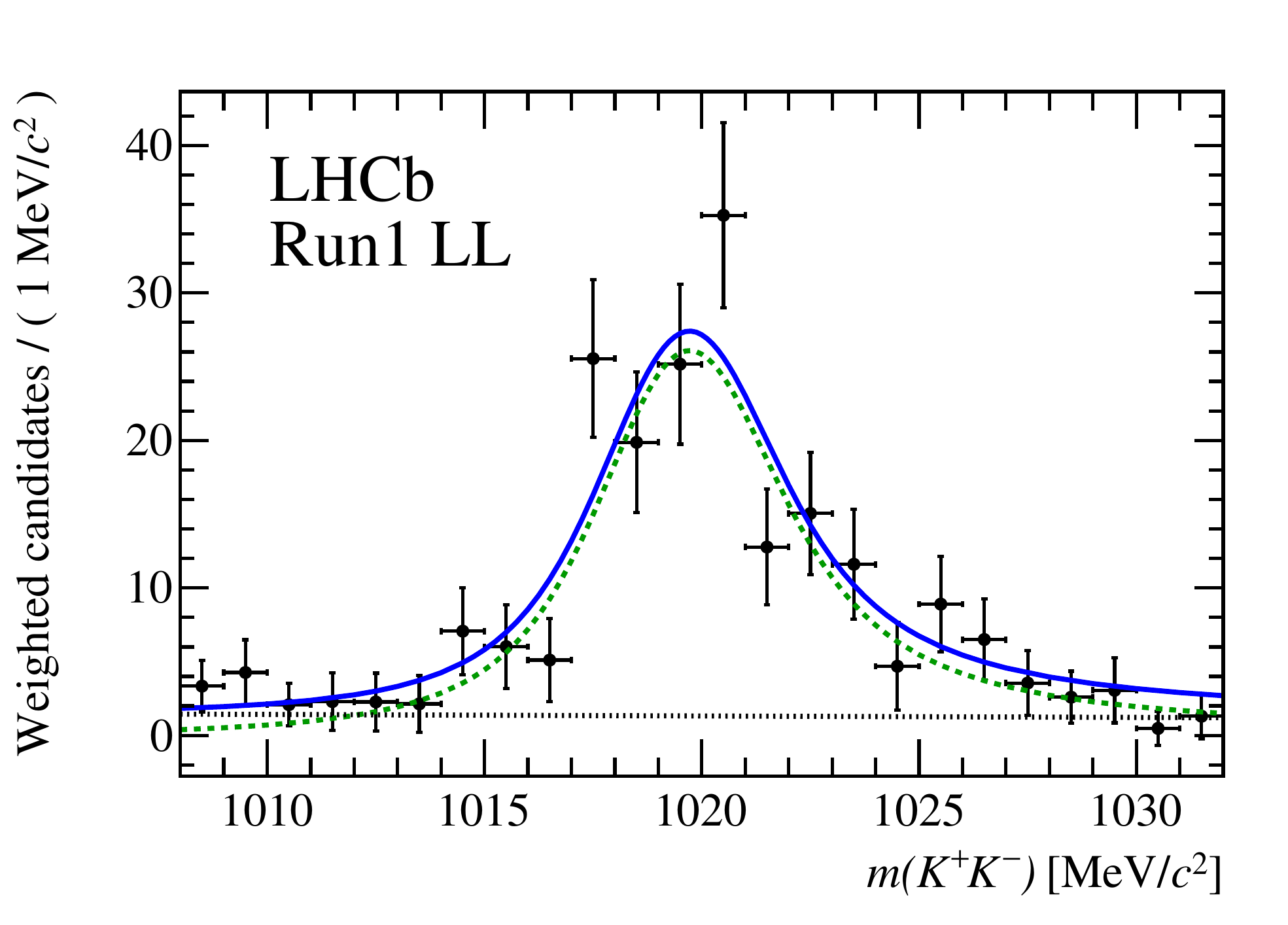}
	\includegraphics[width=0.49\linewidth]{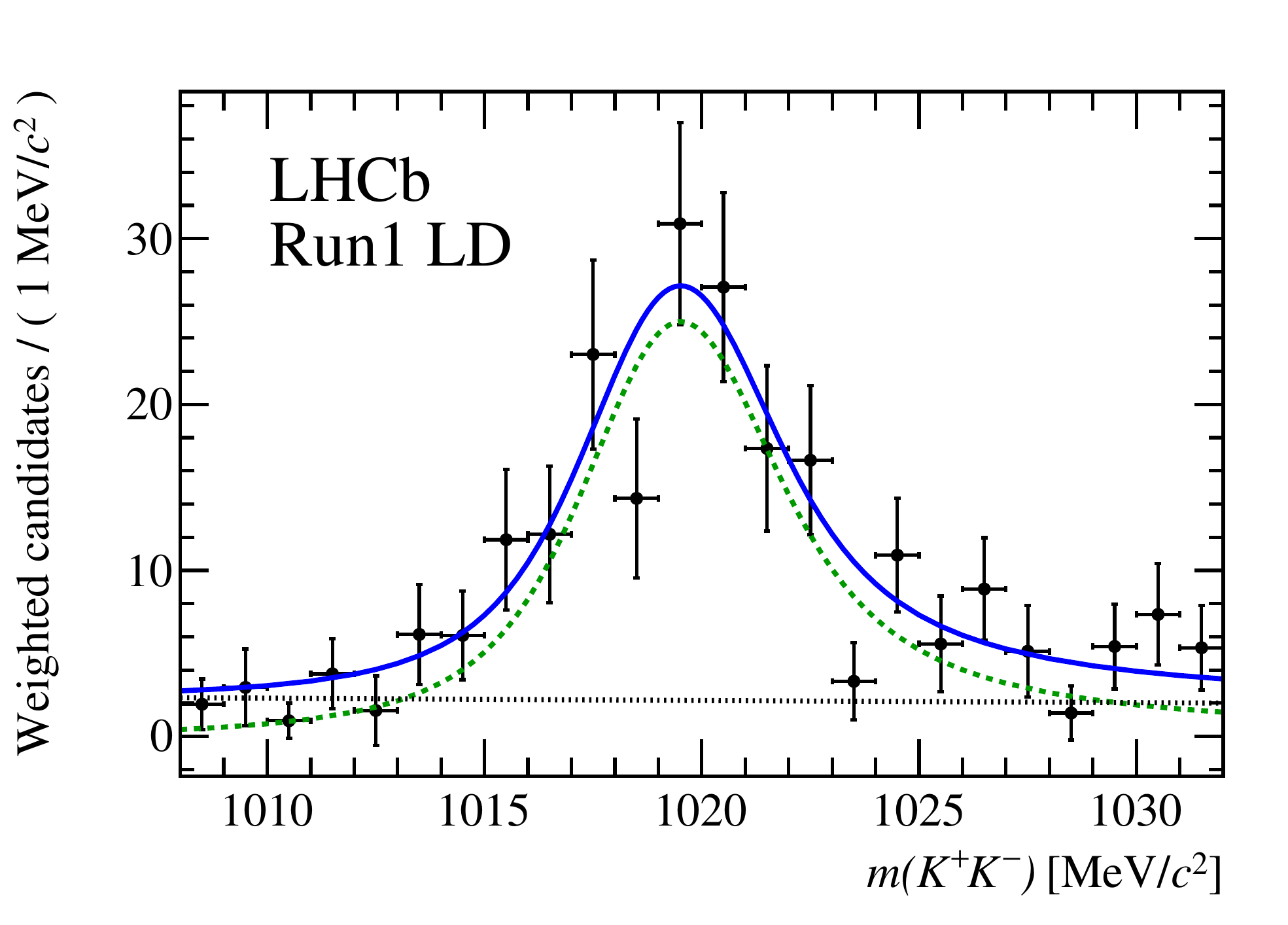}
	\includegraphics[width=0.49\linewidth]{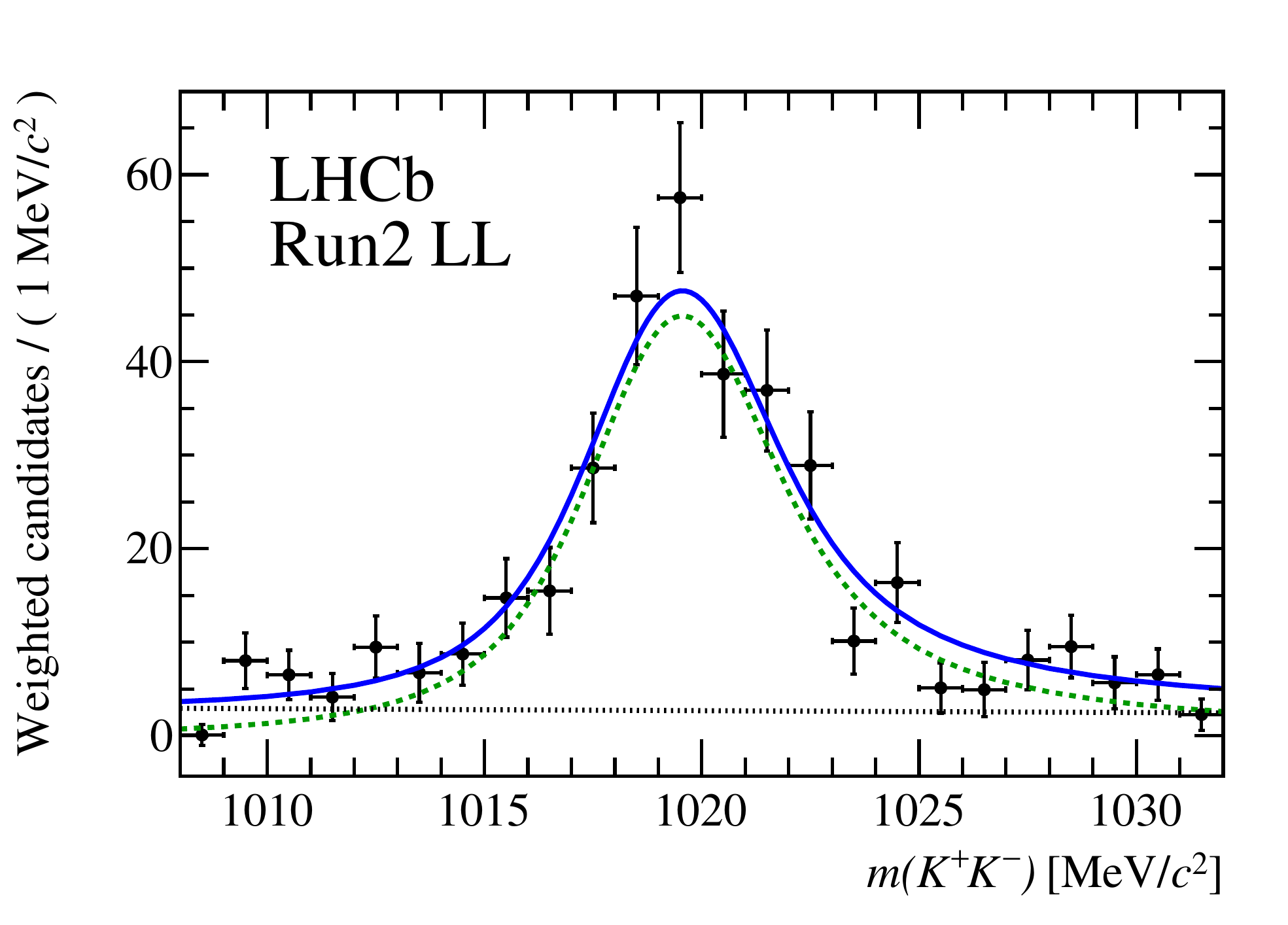}
	\includegraphics[width=0.49\linewidth]{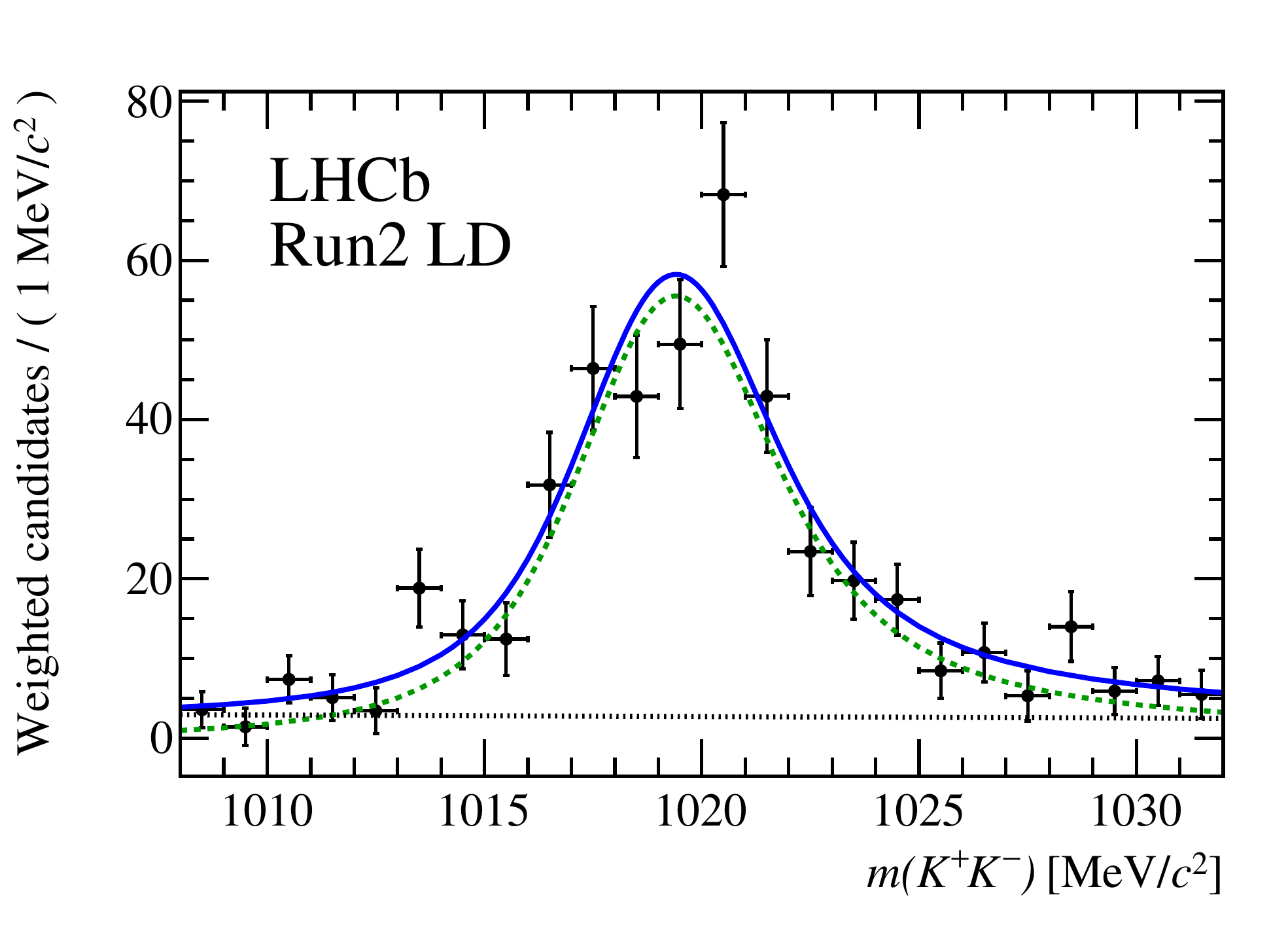}
	\caption{\small Fits to the invariant-mass distribution $m(\Kp\Km)$ of \BzToKSphi data weighted using the \splot technique. The blue curve represents the complete model, the signal \phiz component is given in green (dashed), and the background $f_0(980) \to \Kp \Km$ model is shown in black (dotted).
	}
	\label{fig:massfit_phiinKK}
\end{figure}

Figure \ref{fig:massfit_bd2phiks} shows the $m(\KS\Kp\Km)$ distributions for the Run 1 LL and LD categories. The $m(\Kp\Km)$ distributions for all four data categories are shown in Fig. \ref{fig:massfit_phiinKK}.

\addcontentsline{toc}{section}{References}
 \providecommand{\href}[2]{#2}

\newpage
\centerline
{\large\bf LHCb collaboration}
\begin
{flushleft}
\small
R.~Aaij$^{31}$,
C.~Abell{\'a}n~Beteta$^{49}$,
T.~Ackernley$^{59}$,
B.~Adeva$^{45}$,
M.~Adinolfi$^{53}$,
H.~Afsharnia$^{9}$,
C.A.~Aidala$^{79}$,
S.~Aiola$^{25}$,
Z.~Ajaltouni$^{9}$,
S.~Akar$^{64}$
% ...

{\footnotesize \it
	
% ...
$ ^{9}$Universit{\'e} Clermont Auvergne, CNRS/IN2P3, LPC, Clermont-Ferrand, France\\
% ...
$ ^{25}$INFN Sezione di Milano, Milano, Italy\\
% ...
$ ^{31}$Nikhef National Institute for Subatomic Physics, Amsterdam, Netherlands\\
% ...
$ ^{45}$Instituto Galego de F{\'\i}sica de Altas Enerx{\'\i}as (IGFAE), Universidade de Santiago de Compostela, Santiago de Compostela, Spain\\
% ...
$ ^{49}$Physik-Institut, Universit{\"a}t Z{\"u}rich, Z{\"u}rich, Switzerland\\
% ...
$ ^{53}$H.H. Wills Physics Laboratory, University of Bristol, Bristol, United Kingdom\\
% ...
$ ^{59}$Oliver Lodge Laboratory, University of Liverpool, Liverpool, United Kingdom\\
% ...
$ ^{64}$University of Cincinnati, Cincinnati, OH, United States\\
% ...
$ ^{67}$Syracuse University, Syracuse, NY, United States\\
% ...
$ ^{79}$University of Michigan, Ann Arbor, United States, associated to $^{67}$\\
	\bigskip
% ...
}
\end{flushleft}

\end{document}